\documentclass[12pt]{article}
\usepackage[utf8]{inputenc}
\usepackage[T1]{fontenc}
\usepackage{amsmath}
\usepackage{graphicx}
\usepackage{txfonts}
\usepackage{color}
\usepackage[version=4]{mhchem}
\usepackage{authblk}

\makeatletter
\renewcommand{\@maketitle}{%
  \vspace*{-80pt}
  {\parindent0pt \Large \bfseries \@title \par}
  \vspace{5pt}
  {\parindent0pt \normalsize
  \linespread{1.5}\selectfont 
  \@author \par}
  \vspace{5pt}
  {\footnotesize
  \@date \par} 
}
\renewcommand{\section}{\@startsection{section}{1}{0pt}%
  {-12pt}
  {6pt}
  {\normalfont\normalsize\bfseries}} 
\makeatother

\newenvironment{sciabstract}{%
\begin{quote} \bf}
{\end{quote}}

\newcommand{\spacing}[1]{\renewcommand{\baselinestretch}{#1}\large\normalsize}
\spacing{2}

\title{Wave morphing and flat-top ground states in photonics systems driven by artificial gauge fields}

\author[1,\#]{Peishen Li}
\author[1,\#]{Xiaoyu Zhang}
\author[1]{Feifan Wang}
\author[1]{Ye Chen}
\author[1,*]{Xuefan Yin}
\author[1,2,*]{Chao Peng}
\affil[1]{State Key Laboratory of Photonics and Communication, School of Electronics \& Frontiers Science Center for Nano-optoelectronics, Peking University, Beijing 100871, China}
\affil[2]{Peng Cheng Laboratory, Shenzhen 518055, China}
\affil[\#]{\# These authors contributed equally to this work.}
\affil[*]{To whom correspondence should be addressed; E-mail Xuefan Yin (xuefan\_yin@pku.edu.cn), 
Chao Peng (pengchao@pku.edu.cn).}

\date{}

\begin{document}

\baselineskip 24pt
\maketitle
\clearpage
\begin{sciabstract}
\textcolor{black}{In quantum physics, classical optics, and many other wave systems, wave confinement in a potential well is associated with discrete oscillatory states, and the ground state is typically assumed to vanish uniformly. An open question is whether the ground state can counterintuitively support a flat-top, nonzero envelope, offering new opportunities for quantum emitters, optical antennas, and lasers. Here, we show that by applying the Byers-Yang theorem with an artificial gauge field, energy levels can be continuously shifted, driving eigenstates to morph into a ground state with a uniform yet nontrivial wave envelope. We implement this concept in a photonic crystal slab where a central bulk region is surrounded by heterogeneous bandgaps that engineer reflective phases acting as an artificial local gauge field. By inducing lasing, we probe directly the evolution of the energy levels, demonstrating wave morphing toward a flat-top ground state via near- and far-field measurements.}

\end{sciabstract}

\section*{Introduction}
In a homogeneous, lossless medium, a wave propagates indefinitely without distortion, preserving constant amplitude and a unidirectional flow of energy. When confined within a finite domain, however, boundary-induced reflections lead to counter-propagation and the formation of standing-wave modes, characterized by nonuniform spatial field distributions with distinct nodes and antinodes. Such confinement gives rise to discrete, quantized energy levels $|n\rangle$, a phenomenon common to quantum \cite{bawendi1990quantum,bera2010quantum,jacak2013quantum,lim2015carbon,garcia2021semiconductor}, optical \cite{vaughan2017fabry,marcatili1969bends,braginsky1989quality,armani_ultra-high-q_2003,xu_micrometre-scale_2005}, and acoustic \cite{ingard1953theory, lakin1993high,maccabe2020nano} resonant systems. In these bounded domains, the ground state $|0\rangle$ is typically regarded as trivial, corresponding to a static, zero-amplitude field across the entire region. Consequently, a ground state maintaining a nonzero uniform \textcolor{black}{envelope} and static energy flow is counterintuitive under conventional confinement.
In the literature, waves with flat-top distribution have been surprisingly found in some sophisticated systems, such as photon Bose-Einstein condensates enabled by light-matter interactions \cite{klaers2010bose,pieczarka2024bose, jin2025exciton, schofield2024bose}, photonic crystals (PhCs) engineered to exhibit zero group index\cite{contractor2022scalable, cui2025ultracompacta}, \textcolor{black}{topological systems with non-Hermitian skin effect  \cite{yao2018edge,wang2022non,lin2023topological,lin2024observation}, and gap solitons in nonlinear lattice \cite{Darmanyan1999discrete,2006Soliton,bersch2012optical,goblot2019nonlinear}}. These breakthroughs highlight remarkable opportunities for both fundamental science and emerging technologies, including quantum states with uniform probability distributions over finite regions, \textcolor{black}{self-focusing beam}, and spatially uniform light emission for scalable lasers.

We notice that a particularly straightforward method for realizing a nontrivial ground state is to employ a ring structure with periodic boundary conditions (PBC). Such a geometry naturally supports a ground state $|0\rangle$ with a uniform, nonzero \textcolor{black}{envelope}. According to the Aharonov-Bohm effect \cite {aharonov1959significance, pannetier1984magnetic,tonomura1986evidence,washburn1986aharonovbohm, aronov1987magnetic}, a magnetic flux threading the ring imparts an additional geometric phase to the wavefunction, which in turn shifts the energy spectrum and drives the system toward the ground state $|0\rangle$. This simple yet robust mechanism provides an elegant means of sustaining a flat-top ground state. The underlying principle is encapsulated in the Byers-Yang theorem \cite{byers1961theoretical}, which establishes that the energy levels and their wave profiles vary periodically with the enclosed magnetic flux \cite{gefen1984quantum,gijs1984resistance,chandrasekhar1985observationa,webb1985observationa,morpurgo1998ensembleaveragea,bachtold1999aharonov, peng2010aharonov, rosenow2020fluxa}. Importantly, because the geometric phase is independent of the ring’s specific shape, such flux-induced wave morphing can be realized in any effectively multiply connected system \textcolor{black}{via inducing an artificial gauge field \cite{wu1975concept,pendry2006controlling,dalibard2011colloquium,fang2012realizing,rechtsman2013photonic,aidelsburger2018artificial,lumer2019light,song2025artificial}}.

\textcolor{black}{Here, we propose a general confined photonic wave system, in which the reflective phases at the boundaries act as an artificial gauge field that shifts the energy levels, eventually morphing the energy state to an unusual ground state which maintains a flat-top envelope. Specifically, we encircle a bulk PhC slab with a boundary region, in which counter-propagating waves are connected in a Ouroboros-like configuration that is topologically equivalent to a ring structure, allowing the eigenstates to evolve continuously and periodically with respect to reflective phases. We design and fabricate samples on a III-V/Si hybrid-bonded wafer and realize lasing action to probe the underlying eigenstates. The trajectory of energy-level transitions is tracked, and their asymptotic approaching toward a flat-top ground state is evidenced by observing their distinct lasing patterns.}

\section*{Principle and design}
\begin{figure}[htbp] 
 \centering 
 \includegraphics[width=\textwidth]{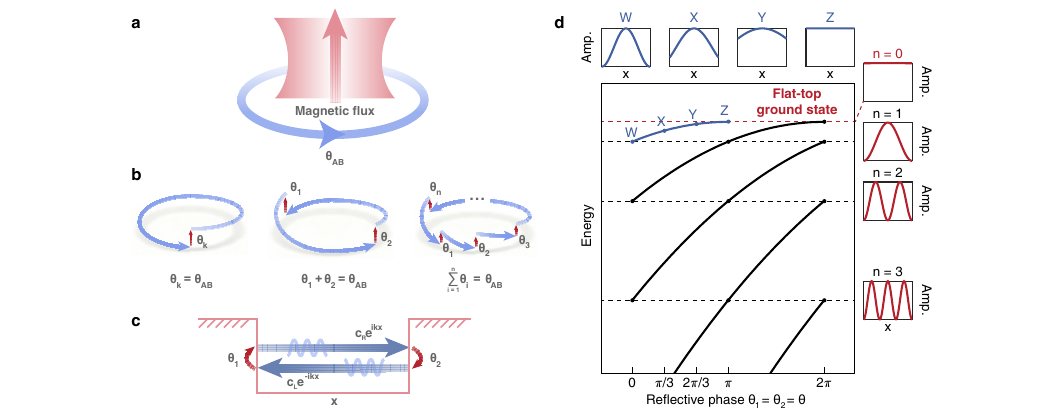}
 \caption{\textbf{Principle of energy level shifting and wave morphing enabled by artificial gauge field.}}
 \label{Fig1}
\end{figure}

We start by considering an artificial ring model (Fig.~\ref{Fig1}a), where a magnetic flux threads a ring of circumference $2L$. Under periodic boundary conditions $\psi(x+2L)=\psi(x)$, the energy spectrum shows as \textcolor{black}{$E_n\propto (n-\theta_{AB}/2\pi)^2/L^2$} according to the Aharonov-Bohm phase $\theta_{AB}$, in which $n$ labels the eigenstate $|n\rangle$ with wavefunction $\psi_n$. The spectrum is periodic in $\theta_{AB}$ by changing the flux, shifting the energy levels $|n\rangle$ even to the ground state $|0\rangle$ as a direct manifestation of the Byers-Yang theorem. Furthermore, with an appropriate gauge choice, the Aharonov-Bohm phase can be embedded into the wavefunctions, and thus the system reduces to \textcolor{black}{a ring with an extra phase kink which imparts $\theta_k=\theta_{AB}$ and modifies the boundary condition to $\psi(x+2L)=\psi(x)e^{-i\theta_k}$ (left panel, Fig.\ref{Fig1}b)}, without altering the spectrum. In general, multiple kinks may be introduced (middle and right panels, Fig.\ref{Fig1}b), with their cumulative phase shift $\theta_{\text{tot}}$ acting as an effective flux. The kink distribution thus functions as a local gauge field: only the total phase is physically relevant, while the rearrangements of the kinks correspond to gauge transformations. Therefore, by appropriately configuring the phase kinks, the ground-state wavefunction $\psi_0$ can be rendered completely static and nontrivially flat \textcolor{black}{(see Supplementary Material Section~1 for details)}. 

Without loss of generality, we discuss a ring model with two phase kinks $\theta_1$ and $\theta_2$ (middle panel, Fig. \ref{Fig1}b), which is equivalent to a potential well (length $L$) bounded by perfectly reflective mirrors with nontrivial phase shifts (Fig. \ref{Fig1}c). In such a system, the wavefunction takes the form $\psi_{1D}=c_R e^{ikx}+c_L e^{-ikx}$ as a superposition of left- and right-propagating waves in a double Ouroboros-like configuration, which yields the boundary conditions of $c_R=-c_L e^{i\theta_1}$ and $c_L=-c_R e^{i\theta_2}e^{2ikL}$ and follows the quantization condition $\theta_{\text{tot}}+2kL=2n\pi$. 
The Byers-Yang theorem still applies: the energy levels periodically shift by varying the total phase $\theta_{\text{tot}}$  (left panel, Fig. \ref{Fig1}d). For simplicity, we set $\theta_1=\theta_2=\theta$ and examine three energy levels $|n=1,2,3\rangle$. Their \textcolor{black}{envelopes} are plotted at an initial phase $\theta=0$ (right panel, Fig. \ref{Fig1}d), showing distinct nodes and antinodes. As $\theta$ increases from $0$ to $\pi$, which makes $\theta_{\text{tot}}$ wind by $2\pi$, all energy levels shift by one cycle, transitioning from $|n\rangle$ to $|n-1\rangle$. Further increasing $\theta$ from $\pi$ to $2\pi$ induces another full cycle, resetting the spectrum to its initial configuration.

Interestingly, in the aforementioned gauge of $\theta=\pi$, the ground state $|0\rangle$ possesses a completely static wavefunction $\psi_0=\text{const}\neq 0$ across the finite-sized potential well. To illustrate this, we track the evolution trajectory from eigenstate $|1\rangle$ to $|0\rangle$ by varying $\theta$ from $0$ to $\pi$ (blue line, Fig.\ref{Fig1}d), and present the corresponding \textcolor{black}{envelopes} at $\theta=0, \pi/3,~2\pi/3,$ and $\pi$, respectively (labeled as $W$-$Z$ in the top panel of Fig. \ref{Fig1}d). As $\theta$ increases, both the wavenumber $k$ and the eigen-energy decrease continuously, and eventually vanish at $\theta=\pi$, \textcolor{black}{accordingly, morphing the wavefunction from being oscillatory to stationary. The wavefunction envelope ultimately becomes completely uniform, thus showing a freezing of energy flow.}

From a mathematical perspective, such an unusual phenomenon can be explained by a homogeneous Neumann condition $(\psi_0'=0)$ at the potential well's boundaries, which naturally supports a ground state $|0\rangle$ with constant, nonzero envelope. Different from the Dirichlet conditions which force the probability density of wavefunctions to vanish at the boundaries, the Neumann conditions only require zero probability flux. This permits the wavefunction to remain finite at the boundaries --- analogous to a string with free ends, allowing full spatial extension of the ground state in a finite domain. At the same time, the vanishing probability flux guarantees no \textcolor{black}{energy} leakage through the boundaries, thereby sustaining wave confinement. \textcolor{black}{Further details are provided in Supplementary Material Section~2}.

\begin{figure}[htbp] 
 \centering 
 \includegraphics[width=\textwidth]{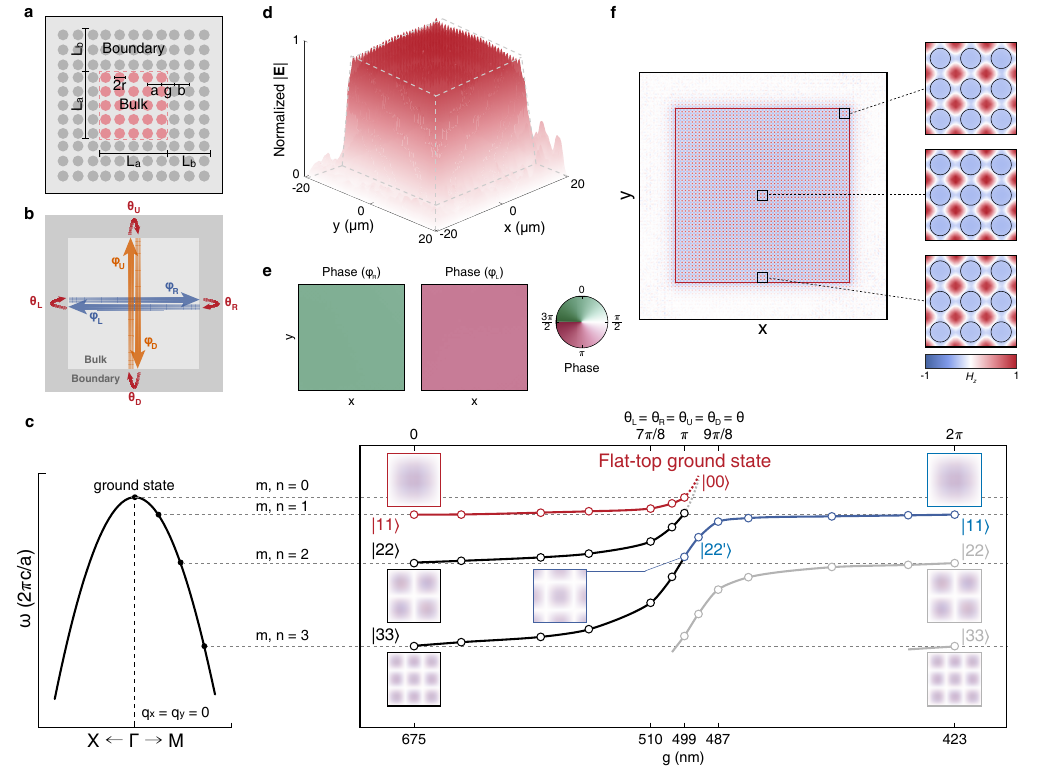}
 \caption{\textbf{A 2D heterogeneous PhC realization.}}
 \label{Fig2}
\end{figure}

To realize a tunable artificial gauge field in a realistic setting, one requires a practical implementation of perfectly reflective boundaries with additional phases. We hereby consider a \textcolor{black}{two-dimensional (2D)} heterogeneous PhC structure (Fig.~\ref{Fig2}a), consisting of a central bulk region enclosed by a surrounding boundary region. \textcolor{black}{Both of the two regions are} patterned as a square lattice of circular air holes whose sizes are given by \textcolor{black}{$L_a$ and $L_b$}, respectively. Their lattice constants $a$ and $b$ are deliberately detuned \textcolor{black}{such that one bulk band} near the second $\Gamma$ point is embedded within the photonic bandgap (PBG) of the boundary region. The reflective phases can be readily controlled by varying the spatial gap $g$ between the two regions. This configuration closely mimics a quantum dot, i.e., a finite-sized confined wave system as we discussed previously \cite{chen2022observation,ren2022low,chen2025observation}.


Next, we show that such a heterogeneous PhC system is equivalent to the artificial ring model presented in Fig. \ref{Fig1}. In the bulk region, the Bloch mode can be approximated as a superposition of two pairs of counter-propagating partial waves $[\phi_Re^{i\beta_0x},~\phi_Le^{-i\beta_0x},~\phi_Ue^{i\beta_0y},~\phi_De^{-i\beta_0y}]$ \cite{liang_three-dimensional_2011,liang2012three,Yang_analytical_2014} as illustrated in Fig.\ref{Fig2}b, where $\beta_0=2\pi/a$ is fast varying wave number associated with the Bloch band. These waves are reflected at the boundaries due to PBG effect, acquiring slow varying envelopes \cite{chen2022observation,chen2025observation,liang2012three}. They can be grouped into two sets as $\phi_{R,~L}$ (blue arrows) and $\phi_{U,~D}$ \textcolor{black}{(yellow arrows)} governed by the reflective boundaries at different directions. \textcolor{black}{Moreover, the waves in one set do not scatter into the other at the boundaries, and thus our 2D system is decomposed into a direct sum of two independent one-dimensional (1D) subsystems.} As an example, we focus on one subsystem of $\phi_{R,~L}(x)=[c_{R,~L}^+\exp{(iq_xx)}+c_{R,~L}^-\exp{(-iq_xx)}]$, where $c_{R,~L}^{+,-}$ are superposition coefficients;  
$q_x$ is slow varying wave numbers, following $q_x\sim \pi/L_a \ll \beta_0$. The reflection between counter-propagating waves forms a closed circulation in which the reflective phases $\theta_{R,~L}$ act as a local gauge field that is topologically equivalent to the ring model in Fig.~\ref{Fig1}. A similar argument holds for $\phi_{U,~D}$, resulting in quantization conditions as:
\begin{eqnarray}
    \label{eq:1}
    2 q_x L_a + \Theta_x(\theta_R+\theta_L) = 2m\pi ~~~~~~~
    2 q_y L_a + \Theta_y(\theta_U+\theta_D) = 2n\pi
\end{eqnarray}
in which $\Theta_{x,y}$ is a monotonic function of $\theta_{R,~L}$ or $\theta_{U,~D}$, acting as the artificial magnetic fluxes. In such a system, the quantized energy levels are denoted as $|mn\rangle$, and  $m, n$ are quantum numbers in $x, y$ directions, respectively. \textcolor{black}{Details are elaborated in Supplementary Material Section~3}.

To quantitatively demonstrate the shifting of energy levels, we numerically solve the eigenstates $|mn\rangle$ by varying the spacing $g$ (Fig. \ref{Fig2}c). For simplicity, we assume $m=n$ and the gap spacings are identical across all directions, which gives $\theta_{R,~L,~U,~D}=\theta$ and the same artificial magnetic flux $\Theta_{x,~y}=\Theta$. The bulk band diagram \textcolor{black}{of a transverse-electric (TE) mode and according} quantized energy levels ($m=n=0,1,2,3$, gray dashed lines) at $\Theta=\theta=0$ are plotted as references (left panel, Fig. \ref{Fig2}c). Note that we use the strength of slow-varying oscillation to present the eigenstates' energy as $E\propto q_x^2+q_y^2$, and thus the state $|00\rangle$ belongs to the exact "zero-energy" ground state, since its momenta align with the band edge as $q_x=q_y=0$. 

The trajectory of energy-level transition is shown in the right panel of Fig.~\ref{Fig2}c. At $\theta=0$ ($g=675$ nm), three eigenstates $|11\rangle, |22\rangle,$ and $|33\rangle$ are identified, whose transverse \textcolor{black}{field distributions} possess distinct oscillations with characteristic \textcolor{black}{antinodes} that define the quantum numbers $m=n\in [1,2,3]$. As $g$ decreases to $499$ nm, the reflective phase increases to $\theta=\pi$ and thus the artificial magnetic flux to $\Theta=2\pi$, causing the entire spectra to shift upward by one level. In particular, state $|11\rangle$ transitions to the ground state $|00\rangle$. \textcolor{black}{As shown in Fig. \ref{Fig2}d, state $|00\rangle$ exhibits a boxcar-like envelope --- uniform within the bulk region owing to $q_{x,~y}=0$, and nearly vanishes in the boundary region. \textcolor{black}{Moreover, although the eigenstate itself can have arbitrary initial phase, all partial waves display uniform phase distributions within the bulk region (Fig. \ref{Fig2}e), and the phase difference between two counter-propagating partial waves (i.e. $\phi_{R,~L}$) remains constant ($\arg{\phi_R/\phi_L}\approx \pi$) throughout the whole bulk. This phase difference is exactly equal to the reflective phase required for $|00\rangle$ ($\theta=\pi$).} Notably, $|00\rangle$ possesses identical field distributions at the center, edge, and corner of the bulk region (Fig. \ref{Fig2}f), indicating an isotropically vanishing group velocity and static energy flow in all lateral directions.} \textcolor{black}{The detailed trajectories can be found in Supplementary material Section~4 and 5.}

Meanwhile, we observe the eigenstate evolution, with $|22\rangle \to |11'\rangle$ and $|33\rangle \to |22'\rangle$. Notably, while $|22\rangle$ at $\theta=0$ and $|22'\rangle$ at $\theta=\pi$ possess the same energy, their field patterns differ by a phase shift. These results indicate that changing the reflective phase from $\theta=0 \to \pi$ ($\Theta=0 \to 2\pi$) does not complete a full evolution cycle. Namely, although the energy levels remain unchanged, their spatial distributions are not restored. Next, as the phase increases from $\theta=\pi \to 2\pi$ by reducing $g$ further from $499$ nm to $423$ nm, $\Theta$ winds by an additional $2\pi$, causing blue-shift of energy levels by one more level. Consequently, at $\theta=2\pi$, the system is restored to its initial configuration at $\theta=0$, both in energy levels and field patterns, thereby completing a full cycle of evolution. In Fig. \ref{Fig2}c, \textcolor{black}{one full cycle is highlighted, consisting of two branches: $|22'\rangle$ to $|11\rangle$ (blue dotted line), and $|11\rangle$ to $|00\rangle$ (red dotted line), respectively.}

Worthy to note that, when $g$ decreases further beyond $499$ nm, the ground state $|00\rangle$ evolves into a special state with frequency above the band edge, denoted as $|\varepsilon\rangle$. This state has effectively ``less energy'' than the ground state $|00\rangle$, and thus can only be described by a non-Hermitian ring model with complex in-plane wavevector $\tilde{q}_{x,y}$. In fact, incorporating non-Hermiticity is also essential when the reflective boundary is imperfect, allowing energy leakage from the bulk \textcolor{black}{into the boundary}. In this case, the system supports a state whose \textcolor{black}{optical field} is mostly localized along the bulk-boundary interface --- a skin-effect resulted from non-Hermiticity. \textcolor{black}{Further details on the non-Hermitian ring model are provided in Supplementary material Section~6}.

\section*{Sample fabrication and experimental setup}
\begin{figure}[htbp] 
 \centering 
 \includegraphics[width=\textwidth]{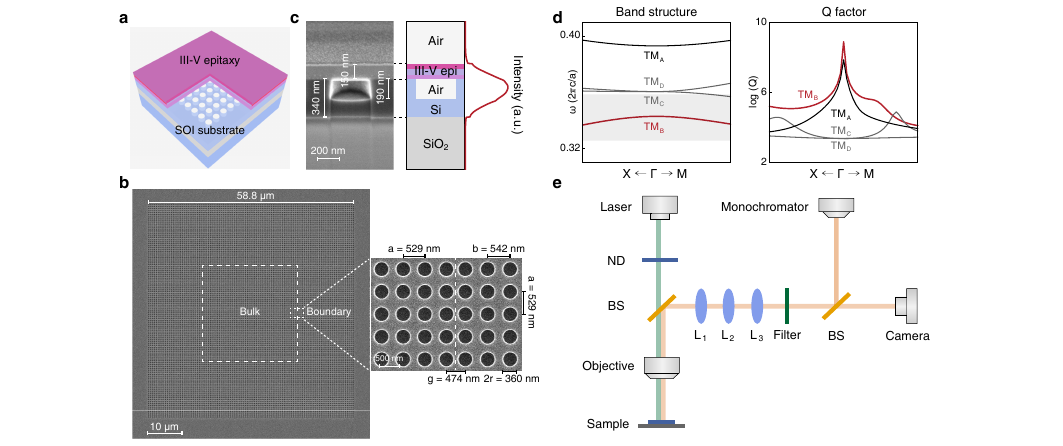}
 \caption{\textbf{Sample design, fabrication and experimental setup.} (a) Schematic of the designed sample on a hetero-bonded wafer, with a SOI patterned as PhCs, and III-V epitaxial layers providing optical gain. (b) Top view of the fabricated PhC (left panel), whose total size is $58.8\times58.8~\mu$m$^2$, and the dashed box distinguishes the bulk region from the boundary region. The right panel shows the zoomed-in details. (c)  Cross-sectional view of bulk PhC obtained by FIB cleaving (left), with its schematic provided for reference (right panel). The red line shows the cross-sectional profile of the TM$_B$ mode as our lasing candidate. (d) Calculated band structures (left) and $Q$ diagram (right panel) of the bulk region. Band TM$_B$ is highlighted in red, and the photoluminescence range is gray-shaded. (e) Schematic of measurements setup, with pumping light (green line) and lasing emission (orange lines). ND: Neutral density filter; BS: Beam splitter. }
 \label{Fig3}
\end{figure}

To experimentally observe the energy level shifting and wave morphing, we employ a \textcolor{black}{hetero-bonded} platform \cite{Zhang:25} consisting of a silicon-on-insulator (SOI) wafer patterned as a PhC \textcolor{black}{slab} and a directly bonded  InP-based epitaxial wafer that provides optical gain (Fig. \ref{Fig3}a). We aim to achieve surface-emitting lasing to characterize the intrinsic properties of \textcolor{black}{eigenstates through observing the radiation from the out-of-plane direction.}

Specifically, fabrication begins with an SOI chip containing a $340$ nm top silicon layer, on which the PhC pattern is defined by electron-beam lithography followed by inductively coupled plasma dry etching. By carefully controlling the etching time, the desired etching depth is achieved. In parallel, an InP-based epitaxial wafer is grown by metal-organic chemical vapor deposition, consisting of four pairs of multi-quantum wells sandwiched between two InGaAsP cladding layers to provide optical gain. The epitaxial wafer is cleaved into chips, then flipped and bonded to the SOI chip via a plasma-assisted bonding process. \textcolor{black}{See Methods for more details about the fabrication process. }

The fabricated sample is examined using scanning electron microscopy (SEM). The top-view image (Fig.~\ref{Fig3}b) shows that the PhC has a total footprint of $58.8\times58.8~\mu\text{m}^2$, comprising both the bulk and boundary regions as designed. A cross-sectional view obtained by focused ion beam (FIB) cleaving (left panel, Fig.~\ref{Fig3}c) reveals the multilayer structure \textcolor{black}{of the bulk region}, with a schematic provided for reference (middle panel). 
We calculate the band structures of the bulk region through three-dimensional (3D) numerical simulations using the measured parameters (Fig.~\ref{Fig3}d). Four transverse-magnetic (TM) modes reside around the second $\Gamma$ point, denoted as TM$_{A,B,C,D}$, among which TM$_{B}$ falls into the material's photoluminescence range. To achieve stable single-mode lasing, we employ shallow etching to realize quasi-mirror symmetry in the vertical geometry. Together with a bound state in the continuum (BIC) \cite{zhen_topological_2014,hsu_bound_2016, wang2024optical} right at the $\Gamma$ point, 
TM$_B$ mode has the relatively highest quality factor \textcolor{black}{($Q$ factor) around the $\Gamma$ point (right panel, Fig.~\ref{Fig3}d)}. Moreover, the cross-sectional optical intensity profile of the TM$_B$ mode is illustrated in the right panel of Fig.~\ref{Fig3}c, indicating that a large portion of energy is confined inside the active layer. Consequently, TM$_B$ is expected to become the dominant lasing mode \textcolor{black}{(see Supplementary material Section~7 for details).}


We use a confocal measurement system to characterize the fabricated sample (Fig.~\ref{Fig3}e). The sample is optically pumped at room temperature with a pulsed $1064$-nm laser (green line). A 10$\times$ objective focuses the pump light onto the sample surface to initiate the out-of -plane lasing (yellow line). Both the lasing beam and reflected pump light are collected by the same objective and relayed through a cascaded $4f$ imaging system (objective + three lenses L$_{1-3}$). A long-pass filter with a cutoff at $1300$ nm removes the residual pump, after which the lasing pattern is recorded by a CMOS camera and the spectrum is measured by a monochromator. Further details of the setup are provided in the Method Section.

\section*{Experimental observation and results}

To observe energy-level shifts and wave-morphing phenomena experimentally, we fabricate six samples $\mathcal{X}\text{,}~\mathcal{Y}\text{,}~\mathcal{Z}\text{,}~\mathcal{U}\text{,}~\mathcal{V}\text{,}~\mathcal{W}\text{,}~$ by using the same fabrication process, \text but with different spacing $g=565,~490,~474,~470,~450,~430$ nm, respectively. All other structural parameters are fixed. The variations in $g$ effectively modify the reflective phases that create \textcolor{black}{an artificial gauge field}. 

\begin{figure}[htbp] 
 \centering 
 \includegraphics[width=\textwidth]{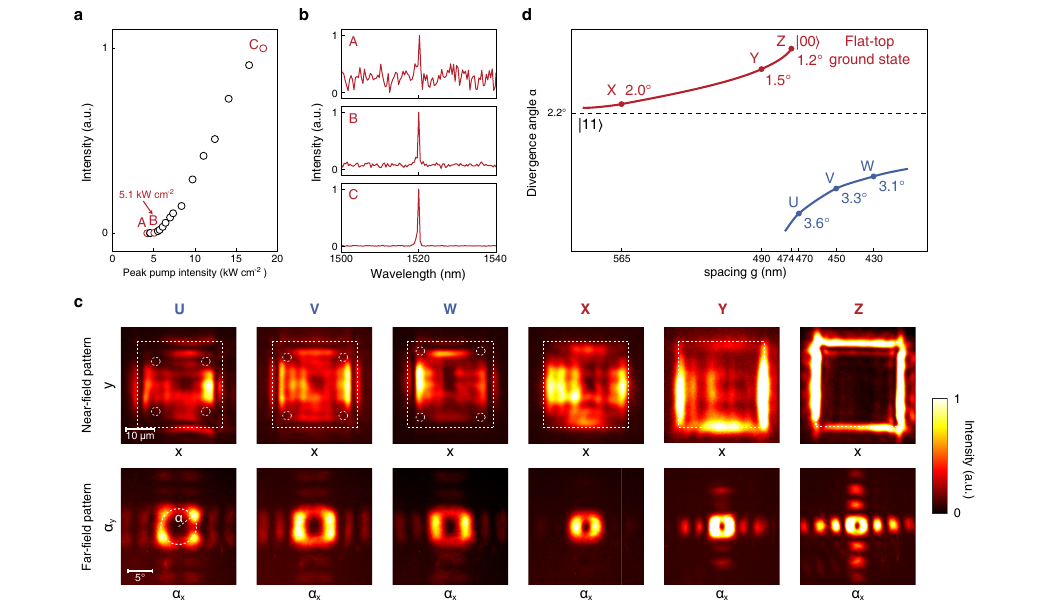}
 \caption{\textbf{Experimental demonstration of energy level shifting and wave morphing.}}
 \label{Fig4}
\end{figure}

First, sample $\mathcal{Z}$ is taken as an example to demonstrate the lasing oscillation (Fig.~\ref{Fig4}a,b). The pump beam is adjusted to cover the entire PhC region, ensuring confinement dominated by the boundary region rather than by gain-guided effects. The power curve in Fig.~\ref{Fig4}a highlights three points ($A$-$C$), with corresponding spectra shown in Fig.~\ref{Fig4}b. At low pump power of $4.2$ kW/cm$^2$ ($A$), only weak resonant feature appears above the spontaneous background. At point $B$, a distinct peak emerges, marking the lasing threshold read as $5.1$ kW/cm$^2$. Further increasing the pump power to $18.3$ kW/cm$^2$ ($C$) yields a sharp single peak, indicating the stable single-mode lasing. \textcolor{black}{Results for other samples are provided in Supplementary material Section~8}. 
Moreover, we measure the near-field pattern (NFP, top panels) and far-field pattern (FFP, bottom panels) for each sample under stable single-mode lasing (Fig.~\ref{Fig4}c). The divergence angle of the lasing beam is further extracted for every sample as a robust and direct indicator of the envelope wavevectors $q_{x,~y}$, enabling quantitative tracking of energy-level shifting.

We begin the tracking with the lasing characteristics of sample $\mathcal{U}$ ($g=470$ nm), designed for $\theta \sim \pi$. The lasing mode is identified \textcolor{black}{to be close to} state $|22'\rangle$, exhibiting a relatively large divergence angle of $3.6^\circ$. In the NFP, four dark spots (white dashed circles) are clearly visible, corresponding to four nodes of its envelope (Fig.~\ref{Fig2}c). Meanwhile, the FFP shows four main lobes with a central dark spot, a typical far-field signature of state $|22'\rangle$ \cite{liang2012three,chen2022observation}. When the spacing $g$ is reduced to $450$ nm ($\mathcal{V}$) and $430$ nm ($\mathcal{W}$), the reflective phase increases beyond $\pi$. In the NFP, the dark nodes pointed by white dashed circles shift toward the PhC \textcolor{black}{corner, while in the FFP, four separate lobes gradually merge into a donut-shaped pattern, with divergence angle decreasing to $3.3^\circ$ ($\mathcal{V}$) and $3.1^\circ$ ($\mathcal{W}$). This observation indicates reducing envelope wavevectors $q_{x,~y}$, agreeing with our theory and confirming that state $|22'\rangle$ is transitioning into $|11\rangle$ when $g$ decreases.} A complete transition to $|11\rangle$ is expected if $g$ is further reduced, but this lies beyond our current fabrication precision.

To demonstrate the wave morphing toward ground state $|00\rangle$, we consider another evolutionary branch from samples $\mathcal{X}$ to $\mathcal{Z}$, where reflective phase evolves from $0$ to $\pi$. \textcolor{black}{In this case, the lasing mode of sample $\mathcal{X}$ ($g=565$ nm) is identified as $|11\rangle$, evidenced by a divergence angle of $2.0^\circ$ that is quite close to the theoretical value ($2.2^\circ$) of $|11\rangle$ state. Also, no dark node appears in its NFP. When the gap is reduced to $490$ nm ($\mathcal{Y}$) and $474$ nm ($\mathcal{Z}$), the donut-shaped FFP continues shrinking to a smaller size, with the divergence angle decreasing to $1.5^\circ$ ($\mathcal{Y}$), and further to only $1.2^\circ$ ($\mathcal{Z}$).} This behavior confirms that the eigenstate is transitioning from $|11\rangle$ towards state $|00\rangle$. At the same time, pronounced side lobes emerge as $g$ decreases, indicating stronger scattering of partial waves at the boundary interface. This trend is clearly visible in the NFP, \textcolor{black}{where the lasing pattern within the bulk region evolves into a smoother distribution, and} the interface region appears significantly brighter.


We claim that the lasing mode in sample $\mathcal{Z}$ ($g=474$ nm) is identified as near-$|00\rangle$ state, with a near-flat-top distribution. As we stated before, there exist a BIC right at the band edge of TM$_B$, whose radiation naturally vanishes. \textcolor{black}{Consequently, when approaching the $|00\rangle$ state, which aligns with the band edge, radiation from the entire bulk region would be strongly suppressed, while scattering-induced emission along the interface dominates. This phenomenon can be verified by lasing characteristics of sample $\mathcal{Z}$: the FFP shrinks to its minimum and the NFP appears almost dark within the bulk region, indicating that the envelope of lasing oscillation is nearly stationary, signifying the appearance of a near-flat-top ground state.}

To acquire a complete view of energy-level shifting, we plot the measured divergence angles of different samples as a function of the spacing $g$, mapping them onto the trajectory of eigenstate evolution (Fig. \ref{Fig4}d). \textcolor{black}{The calculated divergence angle for $|11\rangle$ state ($2.2^\circ$, dashed line) at $\theta=0$ is annotated as a reference.} As $g$ decreases, the reflective phase $\theta$ increases from $0$ to $2\pi$, two distinct evolution branches emerge: $|11\rangle \to |00\rangle$ \textcolor{black}{(samples $\mathcal{X}$-$\mathcal{Y}$-$\mathcal{Z}$, red dotted line)} and $|22'\rangle \to |11\rangle$ \textcolor{black}{(samples $\mathcal{U}$-$\mathcal{V}$-$\mathcal{W}$, blue dotted line). The calculated divergence angle for $|11\rangle$ state acts as the asymptote between the two branches,} agreeing with our theoretical predictions (Fig.~\ref{Fig2}c). \textcolor{black}{In sample $\mathcal{Z}$ with $\theta \sim \pi$, the lasing mode is very close to the ground state $|00\rangle$, exhibiting the minimal divergence angle achievable within our fabrication limits (only $1.2^\circ$).}

\begin{figure}[htbp] 
 \centering 
 \includegraphics[width=9cm]{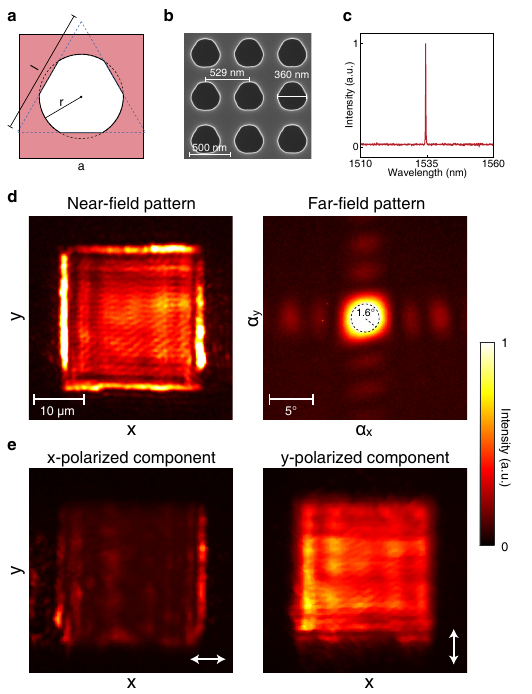}
 \caption{\textbf{Near-flat-top ground state in $C_2$-symmetry-broken system.} (a) Schematic of the irregularly-shaped air hole without $C_2$ symmetry, designed as a circle of radius $r$ cut by an equilateral triangle of side length $l$. (b) SEM image of the fabricated sample with designed irregularly-shaped air holes. (c) Measured single-mode lasing spectrum at a wavelength of $1534$ nm. (d) Measured NFP (left) and FFP (right panel) of the lasing beam. A flat yet nonzero NFP is obtained, arising from the radiation induced by the $C_2$ symmetry breaking. A bright spot in the far field is observed, with a small divergence angle of $1.6^\circ$. \textcolor{black}{(e) Measured $x$- (left) and $y$-polarized (right panel) components of the lasing beam. The $x$-polarized component nearly vanishes while the $y$-polarized component exhibits a flat distribution.}}
 \label{Fig5}
\end{figure}

\textcolor{black}{A particularly intriguing feature of nontrivial ground state $|00\rangle$ is its ability to replicate the band-edge characteristics --- generally considered as a feature of the infinite bulk --- throughout a finite region.} Beyond the system exhibiting a BIC at the $\Gamma$ point, as discussed above, we further demonstrate that this mechanism persists even in a $C_2$-symmetry-broken system. Specifically, we consider irregularly-shaped air holes where the $y$-mirror symmetry is broken (Fig. \ref{Fig5}a). \textcolor{black}{The reflective phases along the $y$-directions must take different values ($\theta_U = 1.08\pi$ and $\theta_D = 0.92\pi$, which still sum to $2\pi$) to realize the near-flat-top ground state $|00\rangle$. } Based on this design, we fabricated a sample (Fig. \ref{Fig5}b) and observed stable single-mode lasing at $1534$ nm (Fig. \ref{Fig5}c).

We measure the NFP and FFP of the lasing mode (Fig. \ref{Fig5}d), where scatterings from the bulk-boundary interface are suppressed by using a square-shaped iris for improved visibility \textcolor{black}{(original data are provided in Supplementary material Section~9)}. Unlike the near-$|00\rangle$ state in the $C_4$-symmetric design (sample $\mathcal{Z}$, Fig. \ref{Fig4}), a flat yet nonzero NFP is obtained, arising from radiation induced by the $C_2$ asymmetry. In the far field, a bright spot rather than a donut-shaped pattern is observed, \textcolor{black}{with a divergence angle of $1.6^\circ$}, \textcolor{black}{which is still smaller than the theoretical value of $|11\rangle$ state ($2.2^\circ$).} \textcolor{black}{Furthermore, as shown in Fig. \ref{Fig5}e, we isolate different polarization components in the NFP by using a linear polarizer. The $x$-polarized component (left panel) is nearly absent, whereas the $y$-polarized component (right panel) remains a pronounced, uniform distribution. This confirms that, under the present boundary reflective phases, the near-$|00\rangle$ state replicates the polarization characteristics of TM$_B$ mode at the $\Gamma$-point --- purely linear-polarization along $y$ direction --- faithfully throughout the bulk region.}

\section*{Conclusions}
\textcolor{black}{We have suggested} that an artificial gauge field can shift energy levels and morph the waves --- a direct manifestation of the Byers-Yang theory, thereby offering a \textcolor{black}{general but} simple method to realize a flat-top ground state, which we have observed in the lasing action of an active PhC slab that is laterally confined by a photonic bandgap with reflective phases. We \textcolor{black}{have} found that this phenomenon is a geometric consequence of the gauge field, rather than a result of the wave interaction details, and is therefore largely insensitive to bulk properties, including geometry, structure size, and wave group velocity. In science, this result unexpectedly reveals a physically feasible route toward a near-ground state, but counterintuitively with a uniform yet nontrivial amplitude. Technologically, this mechanism enables band-edge properties, such as emission strength \cite{jin_topologically_2019,yoshida2023high}, polarization \cite{huang2020ultrafast}, chirality \cite{zhang2022chiral,chen2023observation} and directionality \cite{yin2020observation}, to be replicated over a finite yet, in principle, scalable region, thus opening promising applications from quantum emitters to lasers.


\section*{Acknowledgements}
\textcolor{black}{This work was partly supported by the National Key Research and Development Program of China (2022YFA1404804) and the National Natural Science Foundation of China (62575003, 62135001, 62325501). The simulation was supported by the High-performance Computing Platform of Peking University.}

\section*{Author contributions}
P.L. and X.Z. contributed equally to this work. C.P., X.Y. and P.L. conceived the idea.
X.Y., P.L. and Y.C. performed the theoretical study and simulation. P.L., X.Z. and F.W. conducted the fabrication and measurement. C.P., X.Y. and P.L. wrote the manuscript with input from all authors. C.P. and X.Y. supervised the research. All authors discussed the results.

\section*{Competing interests} The authors declare no competing interests.

\clearpage

\clearpage

\clearpage
\section*{Methods}
\section*{\underline{Numerical simulations}}

Numerical simulations are conducted using the finite-element method implemented in COMSOL Multiphysics. The band structure in the left panel of Fig. 2c is computed using a 2D unit cell with Floquet periodic boundary conditions applied on sides. The evolution and corresponding field distributions of the energy levels shown in the right panel of Fig. 2c are obtained through 2D simulations of the heterogeneous PhC structure. Particular focus is placed on the $|00\rangle$ and $|11\rangle$ states. The field distribution of the $|00\rangle$ state, detailed in Fig. 2d and 2f, is used to derive the phase profiles of $\phi_{R,L}$ via partial wave decomposition (Fig. 2e). The divergence angle of the $|11\rangle$ state in Fig. 4d is determined by applying a Fourier transform to its field profile. The band structures and $Q$ diagram in Fig. 3d are calculated using a 3D unit cell representing the hetero-bonded bulk region, incorporating Floquet periodic boundaries on the sidewalls and perfectly matched layers at the top and bottom surfaces.

\section*{\underline{Sample fabrication}}
The hetero-bonded system comprises two parts: the SOI chip is diced into $10~\text{mm}\times~\text{10mm}$ pieces, consisting of a stack with $340$ nm Si layer, $2~\mu$m SiO$_2$, and a $625\mu$ nm Si substrate; the InP-epi chip is diced into $5~\text{mm}\times5~\text{mm}$ pieces, where the active region consists of four $7.5$ nm compressively strained $In_{0.51}Ga_{0.49}As_{0.932}P_{0.068}$ quantum wells and five $12$ nm lattice-matched $In_{0.75}Ga_{0.25}As_{0.558}P_{0.442}$ quantum barriers, sandwiched by undoped $30$ nm $In_{0.75}Ga_{0.25}As_{0.25}P_{0.75}$ cladding layers, and the thickness of InP substrate is $325~\mu$m. The two parts are bonded via a plasma-assisted bonding process. Firstly, the PhC and VOCs (vertical out-gassing channel) structures are patterned on the SOI chip. During bonding, both the SOI and epitaxial chips undergo through cleaning with organic and inorganic solvents to remove particles and contaminants. An $O_2$ plasma activation step follows, generating hydroxyl-rich functional group on the chip surfaces. The chips are then aligned and pre-bonded at $150C^\circ$ under a $30$ N force for 45 minitues, followed by annealing at $250C^\circ$ under the same force and duration to enhance bonding strength. After bonding, the InP substrate is mechanically ground and polished to a thickness of approximately $35~\mu$m. The remaining InP substrate is then chemically etched using a solution of citric acid, hydrochloric acid, and water in a $1\text{g~:~}8\text{ml~:~}4\text{ml}$ ratio. Finally, the sample undergoes organic cleaning to remove surface particles and is dried with a nitrogen gun.

\section*{\underline{Measurement and data processing}}
We use a $1064$-nm-pulsed laser with a repetition frequency of $10$ kHz and a pulse duration of $2$ ns, to pump our sample at room temperature. As shown in Fig. \ref{Fig3}e, a X10 objective lens (IOPAMI137110X-NIR, Mitutoyo) is used to converge the pump light to a spot and also collect the lasing beam generated from the sample. The following lens L$_1$ ($f$ = $200$ mm) is confocal with the objective lens, as well as L$_2$ ($f$ = $150$ mm) and L$_3$ ($f$ = $300$ mm), forming a $4f$ system. The near-field and far-field images are captured by an InGaAs-infrared-CMOS camera (IMX990, Sony) and undergo data-smoothing with a window size of 9 pixels after deducting CMOS intrinsic noise and ambient background noise. Meanwhile, the lasing beam is focused into a monochromator with a $600g/mm$ optical grating and an infrared array detector (IsoPlane SCT320 and NIRvana 640, Princeton Instruments). 

\section*{Data availiability}
All the data supporting the findings of this study are available in the main paper and its Supplementary Information. Source data are provided with this paper. Additional information can be obtained from the corresponding authors upon request. 

\clearpage
\clearpage
(a)	Schematic of a ring structure threaded by a magnetic flux, which creates a geometric phase $\theta_{AB}$ upon wavefunctions according to the Aharonov-Bohm effect. (b) An effective ring model with different phase kink configurations acting as artificial local gauge fields, including one phase kink $\theta_k$ (left); two phase kinks $\theta_{k1}$ and $\theta_{k2}$ (middle); multiple phase kinks $\theta_{k,i}$ (right). The cumulative phase shift $\theta_{\text{tot}}=\sum_{i=1}^N\theta_{k,i}=\theta_{AB}$ represents an effective magnetic flux. (c) Schematic of a 1D potential well, equivalent to the ring model with two phase kinks (middle panel, b). Two counter-propagating waves ($c_Re^{ikx}$ and $c_Le^{-ikx}$) are reflected by the boundaries with additional phases $\theta_{1,2}$, forming a double Ouroboros-like configuration. (d) Evolutions of energy levels ($|n=1,2,3\rangle$) with respect to reflective phase $\theta=\theta_1=\theta_2$ (left panel). When $\theta$ increases from $0$ to $2\pi$, the effective magnetic flux $\theta_{\text{tot}}=2\theta$ winds by $4\pi$, making the entire set of energy levels shift by two levels, while the eigenstates evolve for a complete cycle. Especially, $|1\rangle$ shifts to the ground state $|0\rangle$ with a flat-top envelope when $\theta=\pi$. The evolution from $|1\rangle$ into $|0\rangle$ is highlighted in blue. The envelopes of four levels ($|n=0,1,2,3\rangle$) at the initial phase $\theta=0$ (right panel), together with the envelopes of eigenstates at $\theta=0, \pi/3, 2\pi/3,$ and $\pi$ on blue trajectory (top panel, labeled $W$–$Z$), are plotted. 

\clearpage
\clearpage
(a)	Schematic of 2D heterogeneous PhC utilized to simulate the artificial local gauge fields, consisting of a bulk region and a boundary region. (b) Schematic of the partial-wave propagating. Four partial waves are grouped into two sets: $\phi_{R,~L}$ (blue arrows) and $\phi_{U,~D}$ (yellow arrows). The reflective phases are denoted as $\theta_{R,~L,~U,~D}$. (c) Band structure of a TE mode (left panel) with gray dashed lines indicating four quantized energy levels ($m=n=0,1,2,3$) at $g=675$ nm. The evolution of energy levels $|m,n=m\rangle$ (right panel) is calculated as $\theta_{R,~L,~U,~D}=\theta$ varies by reducing $g$. When $\theta$ increases from $0$ ($g=675$ nm) to $2\pi$ ($g=423$ nm), a complete evolution cycle is accomplished, as every energy level blue-shifts by two levels. A flat-top ground state $|00\rangle$ is realized at $\theta=\pi$ ($g=499$ nm). The field distributions of energy levels are shown as insets. Evolutions from $|11\rangle \to |00\rangle$ (red dotted) and $|22'\rangle \to |11\rangle$ (blue dotted) are plotted. (d) Envelope of the ground state $|00\rangle$, with the gray dashed lines marking the boxcar-like contour. (e) Phase distribution of partial waves $\phi_{R,~L}$ for $|00\rangle$. (f) Field distribution of $|00\rangle$ in the entire PhC (left panel), with zoomed-in field distributions at its center, edge, and corner (right panels). All results are calculated by using numerical simulation (COMSOL Multiphysics).

\clearpage

\clearpage
\clearpage
(a)	Measured power curve for sample $\mathcal{Z}$ with $g=474$ nm. Three points $A,~B,~C$ are marked with pumping power of $4.2,~5.1,~18.3$ kW/cm$^2$, respectively, showing the status from spontaneous emission, the threshold lasing, to single-mode lasing (wavelength of $1520$ nm).
(b) Measured lasing spectra corresponding to $A$ (top), $B$ (middle) and $C$ (bottom), respectively. (c) Measured 
NFP (top panels) and FFP (bottom panels) of single-mode lasing for six samples with different $g$: $\mathcal{U}$: $470$ nm; $\mathcal{V}$: $450$ nm; $\mathcal{W}$: $430$ nm; $\mathcal{X}$: $565$ nm; $\mathcal{Y}$: $490$ nm; $\mathcal{Z}$: $474$ nm. From the divergence angles, $\mathcal{U}$ ($3.6^\circ$), $\mathcal{X}$ ($2.0^\circ$) and $\mathcal{Z}$ ($1.2^\circ$) are identified as near-$|22'\rangle$, $|11\rangle$ and near-$|00\rangle$ states, respectively. At sample $\mathcal{Z}$ ($g=474$ nm), the FFP shrinks to its minimum (divergence angle of $1.2^\circ$) and the NFP appears almost dark within the bulk region, indicating the achievement of a near-flat-top envelope. In the NFP, the bulk/boundary interface is outlined by a dashed box, and dark nodes on the envelope are highlighted by dashed circles. In the FFP, the divergence angle $\alpha$ is indicated by a dashed circle. (d) Measured divergence angles of different samples as a function of spacing $g$, showing the evolutions of $\mathcal{U}$-$\mathcal{V}$-$\mathcal{W}$ (blue) and $\mathcal{X}$-$\mathcal{Y}$-$\mathcal{Z}$ (red). The dashed gray line represents the theoretical divergence angle for $|11\rangle$ state ($2.2^\circ$).

\clearpage

\end{document}